**A terrain treadmill to study animal locomotion through large obstacles**

Ratan Othayoth[#], Blake Strebel[#], Yuanfeng Han, Evains Francois, *Chen Li

Department of Mechanical Engineering, Johns Hopkins University

*Corresponding author. Email: chen.li@jhu.edu

[#]Equal contributions

**Summary Statement**

A device keeps a small animal on top of a sphere while it traverses obstacles, creating a "terrain treadmill" to study locomotion over a long time and distance at high-resolution.

**Abstract**

A major challenge to understanding locomotion in complex 3-D terrain with large obstacles is to create tools for controlled, systematic lab experiments. Existing terrain arenas only allow observations at small spatiotemporal scales (~10 body length, ~10 stride cycles). Here, we create a terrain treadmill to enable high-resolution observations of animal locomotion through large obstacles over large spatiotemporal scales. An animal moves through modular obstacles on an inner sphere, while a rigidly-attached, concentric, transparent outer sphere rotated with the opposite velocity via closed-loop feedback to keep the animal on top. During sustained locomotion, a discoid cockroach moved through pillar obstacles for 25 minutes (≈ 2500 strides) over 67 m (≈ 1500 body lengths), and was contained within a radius of 4 cm (0.9 body length) for 83% of the duration, even at speeds of up to 10 body length/s. The treadmill enabled observation of diverse locomotor behaviors and quantification of animal-obstacle interaction.

**Introduction**

In nature, terrestrial animals often move through spatially complex, three-dimensional terrain (Dickinson et al., 2000). Small animals are particularly challenged to traverse many obstacles comparable to or even larger than themselves (Kaspari, M , Weiser et al., 1999). By contrast, the majority of laboratory studies of terrestrial locomotion have been performed on flat surfaces (Alexander and Jayes, 1983; Blickhan and Full, 1993; Cavagna et al., 1976; Diederich et al., 2002; Ferris et al., 1998; Full and Tu, 1990; Koditschek et al., 2004; Li et al., 2012; Minetti et al., 2002; Moritz and Farley, 2003; Qian et al., 2015; Spagna et al., 2007; Spence et al., 2010), either rigid or with various surface properties (friction, slope, solid area fraction, stiffness, damping, ability to deform and flow, etc.).

Recent laboratory studies have begun to advance our understanding of animal locomotion in complex terrain with obstacles (Birn-Jeffery and Daley, 2012; Blaesing, 2004; Collins et al., 2013; Daley and Biewener, 2006; Dürr et al., 2018; Gart and Li, 2018; Gart et al., 2018; Harley et al., 2009; Kohlsdorf and Biewener, 2006; Li et al., 2015; Olberding et al., 2012; Parker and McBrayer, 2016; Sponberg and Full, 2008a; Theunissen et al., 2014; Tucker and Mcbrayer, 2012). Because of typical laboratory space constraints, the terrain arenas used in these studies are usually no larger than a few dozen body lengths in each dimension. Thus, they only allow experiments at relatively small spatiotemporal scales beyond ~10 body lengths and ~10 movement cycles. It remains a challenge to study animal locomotion in complex 3-D terrain with large obstacles at larger spatiotemporal scales.

Experiments at large spatiotemporal scales are usually realized by treadmills to keep the animal (including humans) stationary relative to the laboratory (Bélanger et al., 1996; Buchner et al., 1994; Darken et al., 1997; FULL, 1987; Herreid and Full, 1984; Jayakumar et al., 2019; Kram et al., 1998; Leblond et al., 2003; Stolze et al., 1997; Watson and Ritzmann, 1997b; Watson and Ritzmann, 1997a; Weinstein and Full, 1999). However, only small obstacles can be directly mounted on such treadmills (Voloshina et al., 2013); larger obstacles have to be dropped onto the treadmill during locomotion (Park et al., 2015; Snijders et al., 2010; Van Hedel et al., 2002). Furthermore, such linear treadmills allow only untethered movement along one direction. Alternatively, spherical treadmills use lightweight spheres of low inertia suspended on air bearing (kugels) to allow small animals to rotate the spheres as they freely change their movement speed and direction, (Bailey, 2004; Hedrick et al., 2007; Okada and Toh, 2000; Ye et al., 1995). However, the animal is tethered, and obstacles cannot be used.

Here, we create a terrain treadmill (Fig. 1A, B) to enable large spatiotemporal scale, high-resolution observations of small animal locomotion in complex terrain with large obstacles. Our terrain treadmill design was inspired by a celestial globe model. The terrain treadmill consists of a transparent, smooth, hollow, outer sphere rigidly attached to a concentric, solid, inner sphere using a connecting rod (Fig. 1A, Video 1). Terrain modules can be attached to the inner sphere (Fig. 1) to simulate obstacles that small animals encounter in natural terrain (Othayoth et al., 2021). The outer sphere is placed on an actuator system consisting of three actuated omni-directional wheels (Fig. 1A). An overhead camera captures videos of the animal moving on top of the inner sphere, with an ArUCo (Garrido-Jurado et al., 2014) marker attached on its body. The animal's position estimated from tracking the marker is used by a feedback controller (Fig S2C) to actuate the connected spheres with the opposite velocity to keep the animal on top (Fig. 2D, E) as it moves through the obstacle field (Fig. 2A-C, 3A, B, Videos 2, 3). Finally, the reconstructed 3-D motion can be used to estimated different metrics such as body velocities and antennal planar orientation relative to the body heading (Figs. 3, S3, Videos 2, 3).

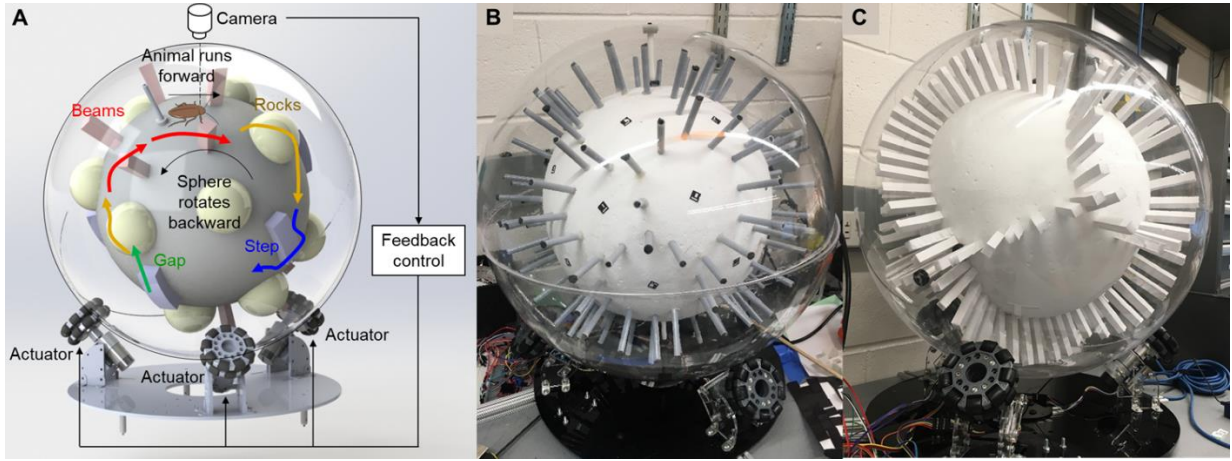

**Fig. 1. Terrain treadmill.** (**A**) Design of terrain treadmill. Colored elements show example modular terrain that can be used. (**B, C**) Terrain treadmill, with (B) sparsely and (C) densely spaced vertical pillars as example terrain modules. ArUCo markers attached on the inner sphere are also shown in (B).

## Results

*Free locomotion at large spatiotemporal scales*

We tested the terrain treadmill's performance in eliciting sustained locomotion of discoid cockroaches ($N$ = 5 animals, $n$ = 12 trials, sparse obstacles) through both sparse (Fig. 1B) and cluttered (Fig. 1C) pillar obstacles (see 'Experimental validation using pillar obstacle field'). Even with cluttered obstacles, where gaps between obstacles were smaller than animal body width, we were able to elicit continuous trials, in which the animal moved through pillars for 25 minutes (≈ 2500 stride cycles) over 67 m (≈ 1500 body lengths) (Video 2). For 83% of the experiment duration, the terrain treadmill contained the animal within a circle of radius 4 cm (0.9 body length) centered about the image center (Fig. 2D, E) even at locomotion speeds of up to 10 body length/s (peak speed of 50 cm/s). We implemented a Kalman filter (Harvey, 1990) to estimate the position of animal and reduce the noise and error in marker tracking (see Supplementary Text). The Kalman filter continued to estimate the animal's position even when the marker was obscured from body rolling (Fig. 2A) or the outer sphere's seam (Videos 1, 2). In addition, over the course of 12 trials, the animal freely explored and visited almost the entire obstacle field (Fig. 3C, D).

Finally, the animal's motion relative to the treadmill was used to estimate metrics such as body velocity components (Fig. 3F-H), antenna planar orientation relative to the body heading (Fig. S3B), and unwrapped 2-D trajectories (Fig. 3E).

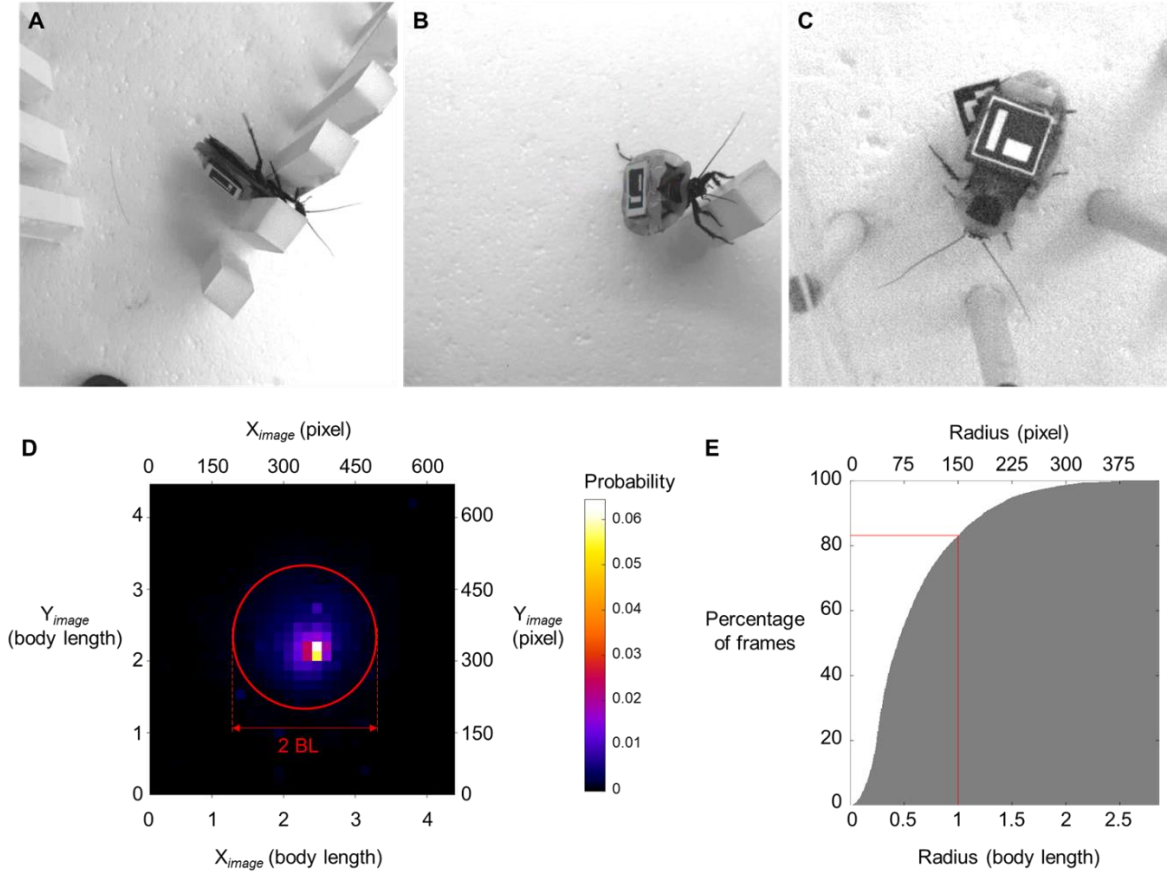

**Fig. 2. Animal behavior and performance of the treadmill.** (**A-C**) Representative snapshots of behaviors including (A) body rolling, (B) body pitching and pillar climbing, and (C) antennal sensing observed during free exploration of terrain. (**D**) Probability of animal's detected location in the image. Red circle of radius 2 animal body lengths is centered at the image center. (**E**) Cumulative histogram of animal's radial position (in body lengths) from the center of the image. Vertical and horizontal red lines show a radius of red circle in (A) and the percentage of frames in which animal's position was maintained within this circle. $N = 5$ animals, $n = 12$ trials.

*Animal-obstacle interaction*

We measured and reconstructed the animal-terrain interaction for 12 trials in which the animal freely explored the sparse obstacle field (Fig. 3, Videos 2, 3). The ArUCo markers attached on the animal and the inner sphere, allowed measuring and reconstructing animal motion relative to obstacle field (see

'Measuring animal movement in obstacle field' in Methods). Because lighting was not optimized, the pillar shadow resulted in substantial variation of the background, and because the left and right antenna are visually similar and often moved rapidly, automated antenna tracking was accurate in only ≈ 40% of frames after rejecting inaccurately tracked data (see 'Automated animal tracking' in Methods). However, this can be improved with refinement of our experimental setup in future (see Discussion).

We then detected which pillar the animal's antennae contacted (Fig. 3 A, B, Video 2) by measuring the minimum distance from each antenna to all nearby pillars. To determine which pillar the antenna interacted with, we determined whether any pillars where within 3 cm from both antennae and which among them were closest to both antennae. We also manually identified the antenna pillar contact, which served as the ground truth. The antenna-pillar contact detected automatically was accurate in over 70% of the contact instances (Fig. S3C).

*Multiple behaviors and behavioral transitions*

In addition to walking or running while freely exploring the obstacle field, the animal displayed other behaviors during interaction with the terrain. For example, when moving in dense obstacle field, the animal often rolled its body in to the narrow gap between the pillars (Fig. 2A) to traverse and occasionally climbed up the pillars (Fig. 2B). In sparse obstacle field, the animal often swept its antennae during free exploration (Fig. 2C, Fig. S3C). The animal also transitioned between these behaviors and occasionally stopped moving (Videos 1, 3).

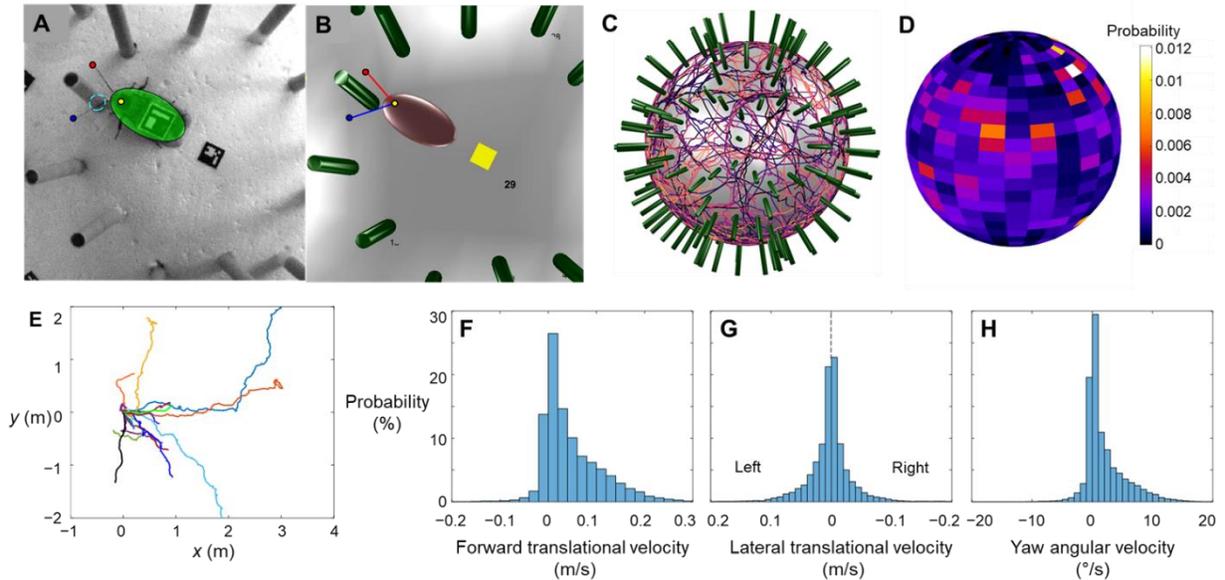

**Fig. 3. Representative metrics and 3-D reconstruction of animal exploring sparse pillar obstacle field.** (**A, B**) Representative snapshot and reconstruction of animal moving through sparse pillar obstacle field.

Transparent green ellipsoid in (A) and brown ellipsoid in (B) show approximated animal body. Red and blue dots show antenna tips. Yellow dot shows the tracked point on animal's head. Dashed cyan circle is the base of the two pillars with which the animal's antenna is interacting. (**C, D**) Ensemble of trajectories (D) and probability density distribution of animal center of mass (E) during free exploration of sparse pillar obstacle field, $N$ = 5 animals, $n$ = 12 trials. (**E**) Unwrapped 2-D trajectories of animal, obtained by integrating forward and lateral translational velocities and yaw angular velocity over duration of trial. (**F-H**) Histogram of animal's (F) forward and (G) lateral translational velocities and (H) yaw angular velocity.

**Discussion**

We created a reconfigurable laboratory platform for large spatiotemporal scale measurements of small animal locomotion through complex terrain with large obstacles (see 'Manufacturing of concentric spheres' and 'Actuation system' in Methods). Compared to existing locomotion arenas, our device increased the limits of experiment duration by ~100× and traversable distance by ~100×. Such large spatiotemporal scales may be useful for studying spatial navigation and memory (Collett et al., 2013; Varga et al., 2017) in terrain with large obstacles, and the larger spatial resolution may be useful for studying interaction of the animal (body, appendages, sensors) with the terrain in detail (Cowan et al., 2006; Dürr and Schilling, 2018; Okada and Toh, 2006). There may also be opportunities to advance neuromechanics of large obstacle traversal by combining the terrain treadmill with miniature wireless data backpacks (Hammond et al., 2016) for studying muscle activation (Sponberg and Full, 2008b) and neural control (Mongeau et al., 2015; Watson et al., 2002). The treadmill design may be scaled down or up to suit animals (or robots) of different sizes. Our treadmill enables large spatiotemporal scale studies of how locomotor behavior emerges from neuromechanical interaction with terrain with large obstacles.

Our study is only a first step and the terrain treadmill can use several improvements in the future to realize its potential. First, we will add more cameras from different views to minimize occlusions and diffused lighting from different directions to minimize shadows, as well as increase camera frame rate to accommodate rapid antenna and body movement, to achieve more reliable tracking of the animal body and antenna through cluttered obstacles during which 3-D body rotations are frequent. Second, feedback control of the sphere can be improved to use not only position but also velocity of the animal to better maintain it on top. This will be particularly useful if the animal suddenly accelerates or decelerates when traversing obstacles. Furthermore, for longer duration experiments, animal could be perturbed when at rest to elicit movement by automatically moving the treadmill. Finally, we need to take into account how locomotion on the spherical treadmill may affect the animal's sensory cues as compared to moving on stationary ground (Buchner et al., 1994; Stolze et al., 1997; Van Ingen Schenau, 1980).

**Methods**

*Manufacturing of concentric spheres*

The hollow, outer sphere was composed of two smooth, acrylic hemispherical shells of radius 30 cm and thickness 0.7 cm (custom ordered from Spring City Lighting, PA, USA; Fig. S1A). The solid, inner sphere was made of Styrofoam (Shape Innovation, GA, USA) and measured 20 cm in radius (Fig. S1A). Both spheres were arranged concentrically using a rigid connecting rod (McMaster Carr, NJ) of diameter 1.25 cm passing through the sphere centers, with a 10 cm space between surfaces of both spheres. To ensure that the connecting rod passed exactly through both sphere centers, we made custom support structures (Fig. S1B, C) to precisely drill through both the inner and outer spheres. The inner sphere was secured to the connecting rod using shaft collars on both sides (Fig. S1A, i). The ends of the connecting rod had threaded holes for the outer hemispheres to be screwed on to it (Fig. S1A, ii). The two outer hemispheres were then mated and sealed using clear tape (3M, MN, USA) without any protrusions to interfere with rotation and with minimal occlusions to the camera's view.

*Actuation system*

The actuation system design followed that of a ballbot (Fankhauser and Gwerder, 2010; Kumaga and Ochiai, 2009; Nagarajan et al., 2014) (but inverted) and consists of three DC motors (Pololu, NV, USA) mounted on a rigid acrylic base (0.6 cm thick, McMaster Carr, NJ, USA), with a set of omni-directional wheels (Fig. S2D-F, Nexus Robots) mounted to each motor. Each set of omni-directional wheel has two parallel wheels which can rotate like a normal wheel about the motor axis. On the rim of each parallel wheel are nine rollers, each of which can rotate about an axis that is perpendicular to the motor axis and tangential to the wheel rim (Fig. S2D). We coated the rollers with a layer of protective rubber (Performix Plasti Dip) to reduce their chance of scratching the transparent outer sphere. The three motors were equally spaced around the base (Fig. S2E) and tilted by 45° (Fig. S2F). The tilt angle was chosen based on the size of the base to allow each omni-directional wheel to be perpendicular to the sphere at the point of contact (Fig. 2F), which reduces vibration and simplifies actuation kinematics. Each DC motor also had an encoder to measure and control its rotation speed and was powered from a 12 V, 30 A DC power supply (Amazon, USA).

*Actuation kinematics*

To measure the relation between the inner/outer spheres rotation and motor rotation, we adapted ballbot's kinematic model (Kumaga and Ochiai, 2009). The desired translational and rotational velocity of the outer sphere's topmost point, and the required motor angular velocities are related as follows:

$$v_{s1} = -v_y \cos\phi - R \sin\phi \, \omega_z \qquad (1)$$

$$v_{s2} = (\frac{\sqrt{3}}{2} v_x + \frac{1}{2} v_y) \cos\phi - R \sin\phi \omega_z \qquad (2)$$

$$v_{s3} = (-\frac{\sqrt{3}}{2} v_x + \frac{1}{2} v_y) \cos\phi - R \sin\phi \omega_z \qquad (3)$$

where $v_{s1}$, $v_{s2}$, $v_{s3}$ are circumferential velocities of the three wheels, $v_x$ and $v_y$ are the fore-aft and lateral velocity components of the outer sphere's highest point, $\omega_z$ is the angular velocity of the outer sphere about the $z$ axis, $\phi$ is the elevation angle of the contact point of each wheel with the outer sphere, and $R$ is the radius of the outer sphere. See Fig. S2A, B for definition of geometric parameters.

*Automated animal tracking*

To track the animal's body and antenna movement, we modified existing automated tracking methods. We attached an ArUCo marker (Garrido-Jurado et al., 2014) on the animal body to track its pose. We chose ArUCo markers because it is feasible for real-time tracking required for fast actuation to keep the animal on top of the treadmill and it can be used to infer 3-D position and orientation using only one camera (Fig. S2G). Prior to each experiment, we adjusted camera position and lens focus to ensure that the topmost point of the inner sphere was in focus. We then calibrated the camera using ArUCo software. Using the calibrated camera, the ArUCo marker on the animal was tracked in real time at 50 Hz.

We used DeepLabCut (Mathis et al., 2018) to track the animal's head and both antennae tips during post processing. We first manually annotated the antenna tips and animal head center in 20 frames each from nine trials, which we used as the sample to train a neural network. The antenna and head positions were then tracked by the trained network (Fig. 3A). We automatically removed some of the obviously incorrect tracking results such as left and right antenna being flipped and obstacles detected as antennae.

*Controlling treadmill motion*

The device used a computer running Robot Operating System (ROS, version: Indigo) (Quigley et al., 2009) to record and track the animal (ArUCo marker), control outer sphere actuation, and collect data (Fig. S2A). We used an overhead camera (PointGrey Flea3) to record animal motion in real time. First, an image was taken by the camera as the animal moved on the inner sphere. This image was used to track the animal's position by detecting the ArUCo marker on animal body. Next, the animal position was filtered using a constant velocity model Kalman filter (Harvey, 1990) (see Supplementary Text), which reduced measurement noise to improve accuracy of animal's estimated position. In addition, the Kalman filter also estimated the animal's position when the animal marker was temporarily occluded by obstacles or was not tracked in real time.

Using the filtered position data, a PID position controller calculated the translational and rotational velocities of the outer sphere's highest point that must be compensated for (control effort) to keep the animal centered on top of the inner sphere (Videos 1, 2). To do so, the controller minimized the error between the position of the marker and the center of the camera viewing area (i.e., the point of the inner sphere directly below the camera's line of sight). We then used the used actuation kinematics (equations (1)-(3)) to determine the motor rotation velocities required to rotate the outer sphere to generate an opposite velocity to that of the animal. Finally, the calculated motor velocities were sent from the computer to an Arduino Due microcontroller, which was used to actuate the motors (via H-bridge motor drivers) to generate the desired rotation. We implemented a PID velocity controller for each motor which measured the motor speed (via the motor encoder) to ensure that desired motor speed was reached.

*Tuning for robust treadmill performance*

Several aspects of the device must be tuned to ensure that the animal remained on top of the inner sphere regardless of its motion relative to the treadmill. First, an appropriate lens focal length and shutter time should be chosen to obtain images with minimal blur for sufficiently fast and reliable marker tracking (Fig. S2G). With the camera placed above 1 m from the top of the inner sphere, we used a 16 mm lens (Fujinon) to obtain a view of sufficient resolution and a 5 ms shutter time to minimize motion blur. We then calibrated the camera using the checkerboard method.

In addition, camera frame rate should be adjusted to not exceed the marker detection rate for a given camera resolution; higher frame rates do not result in better performance if marker detection rate is the bottleneck. Because higher camera resolution increases computational time for marker detection, the smallest resolution that satisfies the other requirements is recommended. In our setup, a resolution of 688 pixels × 700 pixels, marker detection can be performed at ≈50 Hz (Videos 1, 2).

Furthermore, the Kalman filter parameters should be tuned to ensure that the animal's position is tracked sufficiently continuously and smoothly even when the animal accelerates or decelerates suddenly. We found that the most relevant parameter is the noise in the animal's state transition model (i.e., process covariance, see Supplementary Text). Finally, gains of the high-level position PID controller and low-level motor velocity PID controllers should be tuned to track desired treadmill motion as closely as possible while maintaining desired response characteristics such as low overshoot, quick settling time, etc. With tuning, the actuator system can rotate the sphere to achieve desired rotation trajectories accurately (Figs. S2H, Videos 1, 2).

*Experimental validation using pillar obstacle field*

To demonstrate the treadmill's ability to elicit sustained free locomotion of the animal while physically interacting with the terrain, we implemented an obstacle field on the treadmill with tall pillars of a square cross-section of 1.2 cm with gaps between adjacent pillars smaller that the animal body width (Fig. 1C, Video 1). Each rectangular pillar was made of Styrofoam and covered with cardstock on longer faces. We then inserted one end of a toothpick into the pillar, glued them firmly. The other end was then inserted into the Styrofoam inner sphere and the pillar was firmly glued to the inner sphere using hot glue.

Following this, to develop a pipeline for measuring and reconstructing animal's physical interaction with the obstacles, we created an obstacle field with sparsely distributed cylindrical pillars (Figs. 1B, 3C Video 2). Each pillar consisted of a circular plastic tube of height 7 cm and diameter 1 cm, filled with polystyrene foam. To generate an infinitely repeatable obstacle field, we placed the pillars on the inner sphere in a soccer ball pattern. At both ends and midpoint of each edge of the soccer ball pattern, we installed a pillar normal to the spherical surface, with each pillar 4 cm apart from one another. We installed the pillars using technique described above. The supporting rod passing through inner sphere, along with its two shaft collars, also served as two additional pillars of diameter 1.25 cm and height 10 cm, with a cylindrical base of diameter 2.2 cm and height 1 cm.

*Experiment and data collection*

We used discoid cockroaches (*Blaberus discoidalis*) to test the treadmill's ability to elicit free locomotion and measure animal-terrain interaction over large spatiotemporal scales. We put the animal inside the outer sphere and then sealed it. To pick and place the animal onto the inner sphere, we attached a square magnet (16mm side length, 3.5g) on the animal's dorsal side, with an ArUCo marker attached to it for tracking (Fig. S2G, 3A, B). We used a larger magnet to pick up and move the animal to the top of the treadmill and dropped it onto the inner sphere.

We then started the control program to keep the animal on top. The images recorded by the camera were then sent to the ROS program, which first saved each frame in its native format (a bagfile) and then processed the image to track the marker position. Based on the tracked and then filtered marker position, which were used to calculate the velocity of the animal through forward kinematics, motor velocities required to keep the animal centered on were calculated and commanded to the motors. After each experiment, the bagfiles were retrieved and processed using custom MATLAB code to extract the saved images for post processing.

*Measuring animal movement in obstacle field*

To measure the animal's movement relative to the pillar obstacle field, we first measured the movement of the pillar obstacle field (i.e., treadmill rotation) relative to the camera. We attached 31 ArUCo markers to the inner sphere, with one each at the center of hexagonal and pentagonal regions of the soccer ball pattern projected on the sphere (Fig. S1D, S3A). We then separately created a map of all markers attached on the inner sphere (referred to as marker map) using ArUCo Marker-mapper application. Because each marker and its four corners were fixed relative to the coordinate frame attached to the inner sphere (i.e., $T_3$ is known, Fig. S3A), when one of the markers on sphere is tracked (i.e., $T_1$ can be measured, Fig. S3A), the relative pose between sphere body frame and the camera (Fig. S3A, $T_4$) can be computed. When more than one marker on the sphere is detected, relative pose of sphere and camera can be computed by solving the Perspective-*n*-points problem (Lepetit et al., 2009), which estimates camera pose from a known set of 3D points (marker corners) and the corresponding 2D coordinates in the image. The 'solveP*n*P' program in in image processing toolboxes in MATLAB or OpenCV may be used to for this purpose. Because the animal's movement relative to the camera (Fig. S3A, $T_2$) is directly available from tracking via the calibrated camera, the animal's pose relative to the sphere body frame and hence relative to the terrain obstacle field can be calculated (Fig. S3A, $T_5$). Because the ArUCo marker attached to the animal is not necessarily at its center of mass, a constant position and orientation offset must be manually determined and added.

*Unwrapped 2-D trajectory*

Considering that the sphere diameter is ≈ 9× that of animal body length, we approximated the immediate region surrounding the animal to be flat and estimated the animal's equivalent 2-D planar trajectory. To obtain the 2-D trajectory, we integrated the body forward and lateral translational velocities and body yaw angular velocity (Fig. 3F-G) over time, with the initial position at origin and body forward axis along *x* axis. Because during portions of a trial the animal body marker was not tracked for a long duration, we did not consider those video frames. As a result, each trial was assumed to be composed of multiple segments, and each of their equivalent 2-D trajectories were assumed to have the same initial conditions as described above (Fig. 3E).

*Maintenance*

To prevent occlusions and allow reliable camera tracking, the transparent outer sphere must be wiped clean after every use to remove any smudges off the surface. Because wiping with regular cloth towels may scratch the outer sphere, we used a microfiber cloth (AmazonBasics) with soap and water. In addition, we used acrylic cleaner to repair small scratches and dry lubricant (WD-40) to remove tape residue.

**Data Availability**

CAD models, codes for real time control and postprocessing, and data are available at https://github.com/TerradynamicsLab/terrain_treadmill.

**Acknowledgements**

We thank Frank Cook and Rich Middlestadt at the Johns Hopkins University Whiting School of Engineering Manufacturing Facility for assistance with mechanical fabrication, Rafael de la Tijera Obert and Hongtao Wu for installing pillars, and Noah Cowan for discussion.

**Author contributions**

R.O implemented 2-D tracking and 3-D reconstruction, analyzed data, created visualizations, and wrote the paper; B.S. designed and constructed the treadmill, implemented the treadmill control system, and wrote an early draft; Y.H. designed the treadmill and assisted construction; E.F collected animal data for testing treadmill performance; C.L. oversaw the study, designed the treadmill, created visualizations, and wrote the paper.

**Competing interests**

The authors declare no competing or financial interests.

**Funding**

This research was supported by a Beckman Young Investigator award from the Arnold & Mabel Beckman Foundation and The Johns Hopkins Whiting School of Engineering startup funds to C.L., and an NSF Research Experience for Undergraduates (REU) Award in Computational Sensing and Medical Robotics to B.S.

**Supplementary information**

Supplementary information included in submission.

Supplementary Information for article

**A terrain treadmill to study small animal locomotion through large obstacles**

Ratan Othayoth[#], Blake Strebel[#], Yuanfeng Han, Evains Francois, *Chen Li

Department of Mechanical Engineering, Johns Hopkins University

*Corresponding author. Email: chen.li@jhu.edu

[#]Equal contributions

**This PDF contains the following sections**

Supplementary Text

Supplementary Figures S1 – S3

Captions for Videos

**Other supplementary items included with the manuscript**

Videos 1-3

**Supplementary Text**

*Position estimation using Kalman filter*

Kalman Filter enables estimating the state of dynamical system in the presence of noise [67]. We used a constant velocity model Kalman filter to filter the noise from ArUCo marker tracking and estimate the position of the animal relative to the image center. Furthermore, the filter predicts marker position when the marker is not detected due to improper tracking or is temporarily obscured by obstacles. In addition, it suppresses sudden, impulsive rotations of the sphere when tracking is inconsistent. Here, we give a brief overview of the constant velocity Kalman filter implementation. A detailed description is available in [67].

We assume that when the animal is ideally on the topmost point of the sphere, translates in two dimensions (*x* and *y*) and rotates about the *z*-axis (Fig. S2A), with a constant linear velocity ($v_x$ and $v_y$) and angular velocity ($\omega$). In this model, the system state of the animal estimated at a time instance $t-1$ is:

$$\boldsymbol{x}_{t-1} = [x_{t-1} \quad y_{t-1} \quad \theta_{t-1} \quad v_x \quad v_y \quad \omega]^T \tag{1}$$

where $\boldsymbol{x}_{t-1}$ is the system state vector, $(x_{t-1}, y_{t-1})$ are the forward and lateral positions relative to the camera's center point, and $\theta_{t-1}$ is the body yaw of the animal, all at time instance $t-1$. Because the animal's movement is noisy, its system state is not deterministic and associated covariances between each state variable in $\boldsymbol{x}_{t-1}$ at time instance $t-1$ are also defined and represented via a 6 × 6 a state covariance matrix $\mathbf{P}_{t-1}$. We then use the Kalman filter to estimate the future state $\boldsymbol{x}_t$ and covariance $\mathbf{P}_t$ of the animal via two steps: a prediction step followed by a update step.

In the prediction step, we use a dynamic model of how the animal's system state and covariance change due to its motion to obtain an intermediate estimate:

$$\boldsymbol{x}_{t|t-1} = \mathbf{F}_t \boldsymbol{x}_{t-1} \tag{2}$$

$$\mathbf{P}_{t|t-1} = \mathbf{F}_t \mathbf{P}_{t-1} \mathbf{F}_t^T + \mathbf{Q}_t \tag{3}$$

where $\boldsymbol{x}_{t|t-1}$ and $\boldsymbol{P}_{t|t-1}$ are the intermediate estimates of state variables and covariances, $\mathbf{F}_t$ is the state transition matrix at time instance $t$ (see Eqn. (8) for definition), $\mathbf{Q}_t$ is the process covariance matrix at time instance $t$ (see Eqn. (9) for definition). Becaues $\mathbf{F}_t$ and $\mathbf{Q}_t$ may not model the changes in state and state covariances exactly, we then measure the system output to reduce the error in measurements:

$$\boldsymbol{y}_t = \boldsymbol{z}_t - \mathbf{H}_t \boldsymbol{x}_{t|t-1} \tag{4}$$

where $\boldsymbol{y}_t$ is the difference between measured system output $\boldsymbol{z}_t$ and expected system output ($\mathbf{H}_t \boldsymbol{x}_{t|t-1}$) and $\mathbf{H}_t$ is the observation model that maps the system state into the expected system output (see Eqn. (10) for definition). Next, we calculate the Kalman gain matrix:

$$\mathbf{K}_t = \mathbf{P}_{t|t-1} \mathbf{H}_t (\mathbf{H}_t \mathbf{P}_{t|t-1} \mathbf{H}_t^T + \mathbf{R}_t)^{-1} \tag{5}$$

where $\mathbf{K}_t$ is the Kalman gain matrix, $\mathbf{R}_t$ is the covariance of the noise in measurement $\boldsymbol{z}_t$ (see Equation 11 for definition).

Using the calculated Kalman gain, we then perform the update step in which we update the system state and covariances using as follows:

$$\boldsymbol{x}_k = \boldsymbol{x}_{t|t-1} - \mathbf{K}_t \boldsymbol{y}_{t|t-1} \tag{6}$$

$$\mathbf{P}_t = (\mathbf{I} - \mathbf{K}_t \mathbf{H}_t)\mathbf{P}_{t|t-1} \qquad (7)$$

where $x_t$ and $\mathbf{P}_t$ are the system state and covariance matrix at time instance $t$ and $\mathbf{I}$ is the identity matrix of appropriate dimension.

For our constant velocity model, $\mathbf{F}_t$, $\mathbf{Q}_t$, $\mathbf{H}_t$, and $\mathbf{R}_t$ are as follows:

$$\mathbf{F}_t = \begin{bmatrix} 1 & 0 & 0 & \Delta t & 0 & 0 \\ 0 & 1 & 0 & 0 & \Delta t & 0 \\ 0 & 0 & 1 & 0 & 0 & \Delta t \\ 0 & 0 & 0 & 1 & 0 & 0 \\ 0 & 0 & 0 & 0 & 1 & 0 \\ 0 & 0 & 0 & 0 & 0 & 1 \end{bmatrix} \qquad (8)$$

$$\mathbf{Q}_t = 0.05 \times \begin{bmatrix} \frac{\Delta t^4}{4} & 0 & 0 & \frac{\Delta t^2}{2} & 0 & 0 \\ 0 & \frac{\Delta t^4}{4} & 0 & 0 & \frac{\Delta t^2}{2} & 0 \\ 0 & 0 & \frac{\Delta t^4}{4} & 0 & 0 & \frac{\Delta t^2}{2} \\ \frac{\Delta t^2}{2} & 0 & 0 & \frac{\Delta t^2}{2} & 0 & 0 \\ 0 & \frac{\Delta t^2}{2} & 0 & 0 & \frac{\Delta t^2}{2} & 0 \\ 0 & 0 & \frac{\Delta t^2}{2} & 0 & 0 & \frac{\Delta t^2}{2} \end{bmatrix} \qquad (9)$$

$$\mathbf{H}_t = \begin{bmatrix} 1 & 0 & 0 & 0 & 0 & 0 \\ 0 & 1 & 0 & 0 & 0 & 0 \\ 0 & 0 & 1 & 0 & 0 & 0 \end{bmatrix} \qquad (10)$$

$$\mathbf{R}_t = \begin{bmatrix} 0.00001 & 0 & 0 \\ 0 & 0.00001 & 0 \\ 0 & 0 & 0.00001 \end{bmatrix} \qquad (11)$$

where $\Delta t$ is the time step between two estimates.

**Supplementary Figures**

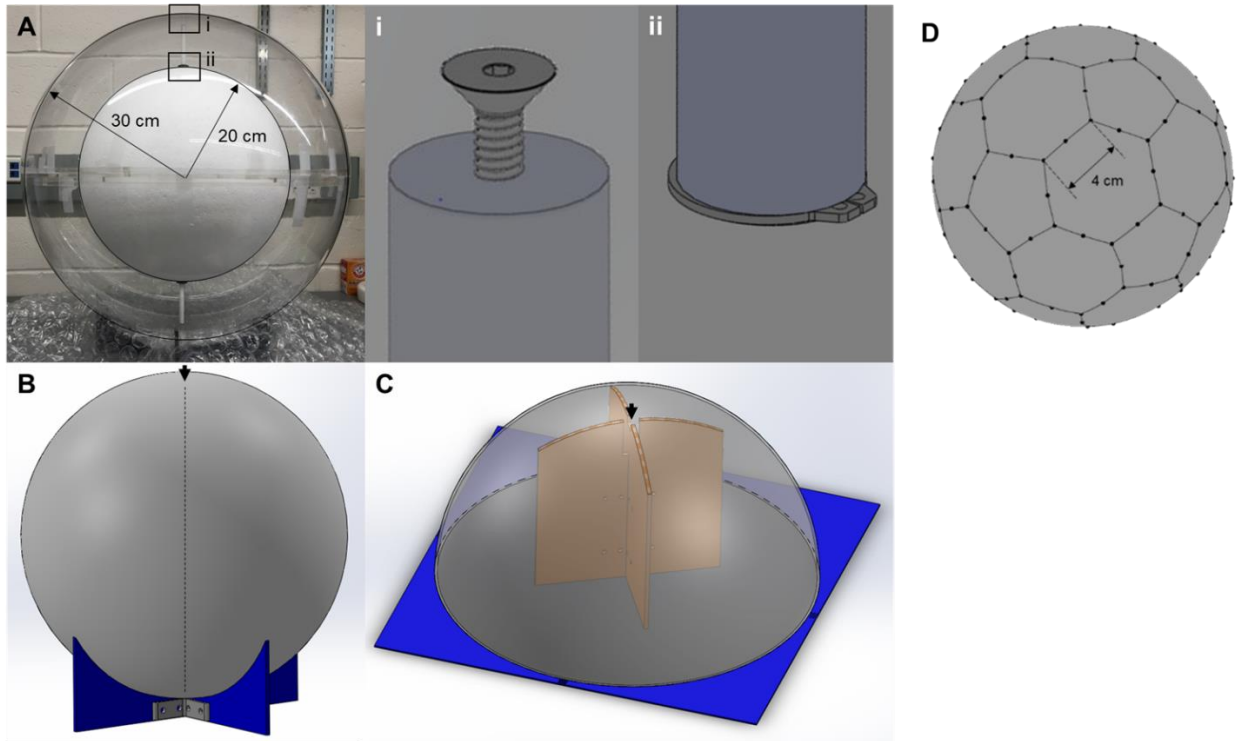

**Fig. S1. Manufacturing of concentric spheres of terrain treadmill.** (**A**) Photo of rigidly attached, concentric spheres. Insets show fasteners that secure inner (i) and outer (ii) spheres to connecting rod. (**B**) Support for drilling inner sphere. (**C**) Support for drilling outer sphere. Arrows in (B, C) show locations of drilling. (**D**) Soccer ball pattern projected onto the inner sphere. Black dots show location of pillars.

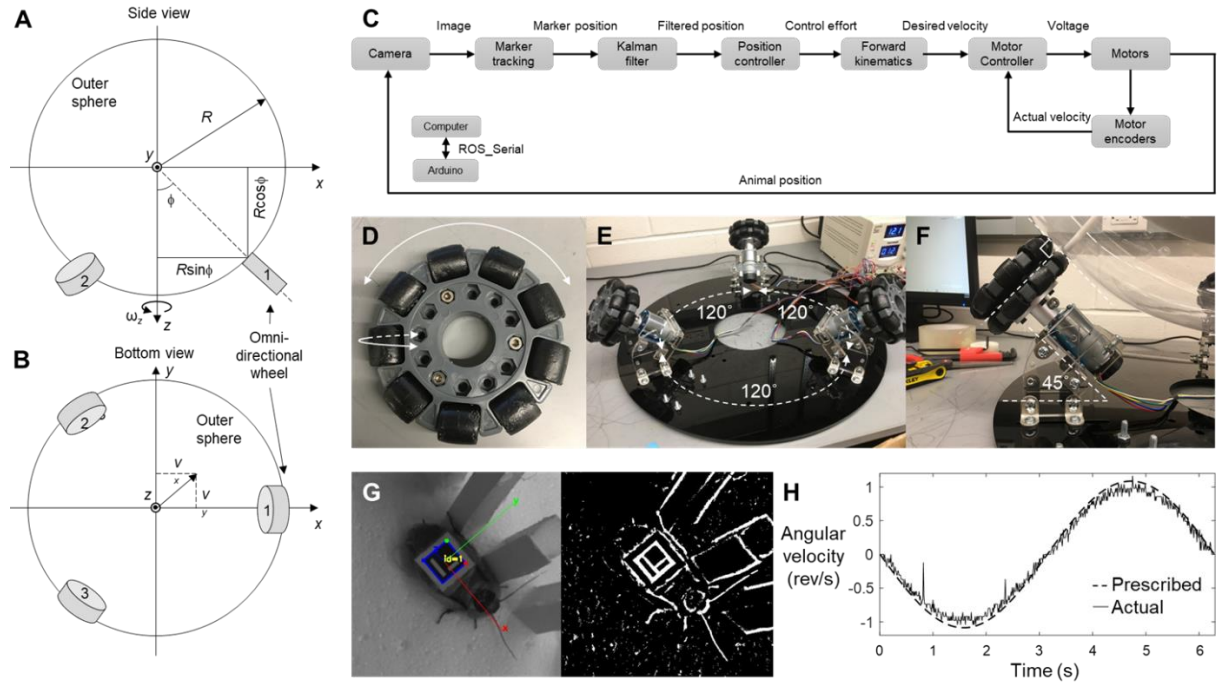

**Fig. S2. Overview of treadmill design, actuation, and control.** (**A-B**) Definition of geometric parameters for forward kinematics. (**A**) Side view. (**B**) Bottom view. Gray discs are omni-directional wheels and white circle is outer sphere. (**C**) Block diagram of treadmill's control system. (**D**) Omni-directional wheel. Large arrow shows rotation of the entire wheel; small arrow shows rotation of the small roller. (**E**) Treadmill actuator system consisting of omni-wheels mounted on DC motors. Each of the three circularly arranged actuators are 120° apart. (**F**) Inclination of motor-omni-wheel assembly relative to the base. Omni-wheel of each actuator is perpendicular to the transparent outer sphere. (**G**) Automated tracking of animal position using an ArUCo marker. Left: visible light camera view. Right: extracted outline using image processing. (**H**). Comparison of prescribed (dashed) and actual (solid) angular velocity of the sphere as a function of time during a simple rotation about a fixed axis.

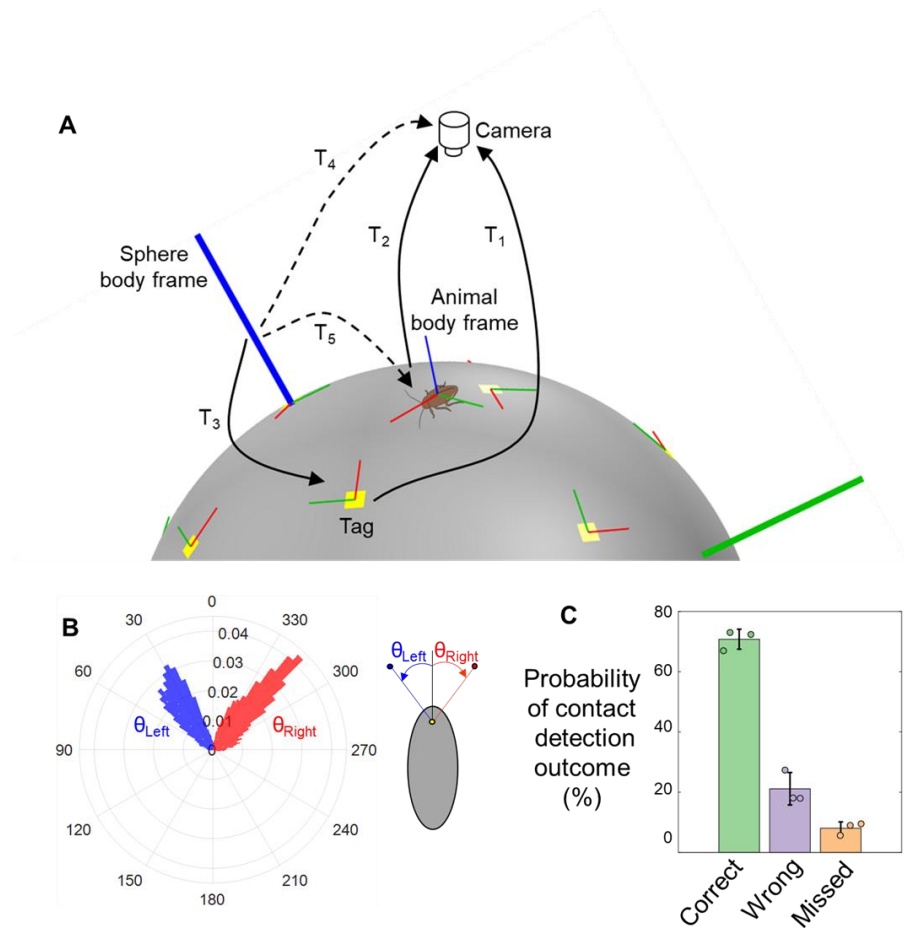

**Fig S3. Measuring motion of animal exploring sparse pillar fields.** (**A**) Coordinate frame transformation to measure animal motion relative to sphere. Solid black arrows are relative 3-D poses ($T_1$, $T_2$ and $T_3$) that are known or measured directly from acquired images. Dashed arrows are the two relative 3-D poses ($T_4$ and $T_5$) that are calculated from measurements to obtain animal motion relative to the sphere. Yellow squares with red and green lines show the markers attached to the sphere and their *x* and *y* axes, respectively. Thick green and blue lines show the *y* and *z* axes of the frame attached to the inner sphere. (**B**) Histogram of left ($\theta_{left}$, blue) and right ($\theta_{right}$, red) antenna planar orientation relative to body heading (see schematic on right for definition). (**C**) Accuracy of antenna-pillar contact detection outcomes. $N = 3$ animals, $n = 3$ trials.

**Captions for Videos**

[Video 1.](#) **Animal exploring terrain treadmill.** Left: Animal maintained on top of the treadmill while traversing a sparse pillar field. Right: A complete, continuous trial of 25 minutes duration. Played at 50× speed.

[Video 2.](#) **Reconstructed 3-D motion of the animal.** Overhead view of animal traversing a sparse pillar field (left) and its 3-D reconstruction (right). Transparent green ellipsoid (left) and brown ellipsoid in (right) show approximated animal body. Red and blue dots show tracked antenna tips. Yellow dot shows tracked point on head. Dashed cyan circle (left) is the base of the detected pillars with which animal antenna is interacting.

[Video 3.](#) **Representative behaviors trajectories of the animal on the treadmill through a pillar field.** Left: Different behaviors of the animal. Right: Reconstructed trajectories and 3-D motion of animal traversing pillar obstacle field in reprensentative trials. Green ellipsoid shows approximated animal body. Sudden jumps in reconstruction are an artifact of manually omitting sections of trial when animal is not tracked accurately.